\documentclass[aps,twocolumn,prl]{revtex4}
\usepackage{epsfig}
\usepackage{graphicx}
\usepackage{amssymb,amsmath,bbold}
\usepackage{moreverb}

\newcommand {\be}{\begin{equation}}
\newcommand {\ee}{\end{equation}}
\newcommand {\ba}{\begin{eqnarray}}
\newcommand {\ea}{\end{eqnarray}}

\begin{document}
\title{Macroscopic Quantum Tunneling of Solitons in Bose-Einstein Condensates}
\author{J. A. Glick and Lincoln D. Carr}
\affiliation{Department of Physics, Colorado School of Mines, Golden, CO 80401, USA}
\date{May 25, 2011}

\begin{abstract}
We study the quantum tunneling dynamics of many-body entangled solitons composed of ultracold bosonic gases in 1D optical lattices. A bright soliton, confined by a potential barrier, is allowed to tunnel out of confinement by reducing the barrier width and for varying strengths of attractive interactions. Simulation of the Bose Hubbard Hamiltonian is performed with time-evolving block decimation. We find the characteristic $1/e$ time for the escape of the soliton, substantially different from the mean field prediction, and address how many-body effects like quantum fluctuations, entanglement, and nonlocal correlations affect macroscopic quantum tunneling; number fluctuations and second order correlations are suggested as experimental signatures. We find that while the escape time scales exponentially in the interactions, the time at which both the von Neumann entanglement entropy and the slope of number fluctuations is maximized scale only linearly.
\end{abstract}

\maketitle
Tunneling is one of the most pervasive concepts in quantum mechanics and is essential to contexts as diverse as biophysics~\cite{liYF2010}, the $\alpha$-decay of nuclei, vacuum states in quantum cosmology~\cite{coleman1977} and chromodynamics~\cite{ostrovsky2002}. Macroscopic quantum tunneling (MQT), the aggregate tunneling behavior of a quantum many-body wavefunction, has been demonstrated in many condensed matter systems~\cite{blencoweM2004} and is one of the remarkable features of Bose-Einstein Condensates (BECs). For example, predictions for MQT in BECs range from cold atom Josephson rings~\cite{solenov2010} to collapsing BECs~\cite{ueda1998}, and MQT has been observed in double well potentials~\cite{zapata1998}. MQT has mainly been treated under semiclassical approximations such as JWKB and instanton methods, while more recently significant progress has been made towards a more general many-body picture via multi-configurational Hartree-Fock theory~\cite{streltsov2009}. We present the first fully many-body entangled dynamical study of the quantum tunneling escape problem.

In this Letter, we study how an initially trapped bright soliton composed of an ultracold bosonic gas in an optical lattice escapes via quantum tunneling through a barrier.  In particular, we find that the maximum of both the von Neumann entanglement entropy and the slope of the number fluctuations between the initially trapped region behind the barrier and the escape region increases linearly with the interaction strength, while the \emph{escape time} $t_{\mathrm{esc}}$, i.e., the time at which the average number of remaining particles falls to $1/e$ of its initial value, increases exponentially with interactions. The optical lattice is required to use time-evolving block decimation (TEBD)~\cite{vidal2004}, and also serves to increase quantum depletion and drive the system towards the strongly non-semi-classical regime.

In the mean field limit, the macroscopic wavefunction of the condensate is well-described by the Gross-Pitaevskii or nonlinear Schr\"{o}dinger equation (NLS). The quantum many-body escape problem has already been studied in this context, where it has been found to have quite different features from single-particle quantum tunneling, including a tunneling time which is not simply the inverse of the JWKB tunneling rate~\cite{carr2005c} and a non-smooth dynamical behavior referred to as ``blips,'' in which particles escape through the barrier in bursts~\cite{dekel2010}.

The NLS admits bright and dark solitonic solutions; we focus on the case of attractive interactions, of strong interest in recent experiments due to the newly demonstrated ability to change the interaction sign and strength over seven orders of magnitude~\cite{pollack2009}.  Solitons are localized, robust waves that propagate over long distances without changing shape, due to the nonlinearity in the NLS, which counteracts dispersion.  However, away from the mean field limit, quantum fluctuations affect the dynamics and stability of solitons~\cite{carr2009l}.  Bright solitons in BECs have already been observed~\cite{carr2002b}, including in optical lattices~\cite{eiermann2004}, so that our predictions and ideas can be tested with present experimental apparatus.  In particular, the small number of atoms we work with, from a few to 70, can be created in a 2D array of 1D systems~\cite{kinoshita2006}.  Proposed applications of bright solitons include a pulsed atomic laser~\cite{carr2004h}. In such applications, MQT, quantum entanglement, and quantum fluctuations are all critical contributors to the overall dynamics and stability of solitons, and must be taken into account.

Consider a system of $N$ bosons at zero temperature in the canonical ensemble held in a 1D homogeneous lattice of $L$ sites, with box boundary conditions, similar to Ref.~\cite{kinoshita2006}. The lattice is sufficiently deep to allow us to evoke a lowest Bloch band tight-binding approximation. Then the system is well-described by a discretized version of the full many-body Hamiltonian, i.e., the Bose Hubbard Hamiltonian (BHH),
\begin{equation}
\label{eq:BHH}
\hat{H} = -J\sum_{i=1}^{L-1}(\hat{b}_{i+1}^\dagger\hat{b}_i+\mathrm{h.c.})+\sum_{i=1}^L [\frac{U}{2}\hat{n}_i(\hat{n}_i-\hat{\mathbb{1}})+\hat{V}^{\mathrm{ext}}_i].
\end{equation}
In Eq.~\eqref{eq:BHH}, $J$ is the energy of hopping and $U<0$ determines the on-site two-particle interactions. An external rectangular potential barrier, of width $w$ and height $h$, is given by $\hat{V}^{\mathrm{ext}}$. The field operator $\hat{b}_{i}^\dagger$ ($\hat{b}_{i}$) creates (annihilates) a boson at the $i\mathrm{th}$ site and $\hat{n}_{i} \equiv \hat{b}_{i}^\dagger \hat{b}_{i}$. We will work in hopping units: $J=1$ and time $t$ in units of $\hbar/J$.

To evolve the BHH in real and imaginary time we use TEBD~\cite{vidal2004}. TEBD is a matrix product state numerical method that time evolves Eq.~\eqref{eq:BHH} on an adaptively reduced Hilbert space, given that the system is lowly entangled.
TEBD is necessary because MQT is triggered by long range density fluctuations, and thus poorly modeled by mean field theory~\cite{takagi1997}. Instanton methods offer another approach towards calculating tunneling rates within a semiclassical approximation~\cite{vainstein1982}, but are rendered inaccurate for larger interaction strengths~\cite{danshita2010}, whereas TEBD suffers from no such limitations.

To describe the system from a mean field perspective, the discrete NLS (DNLS) may either be obtained via discretization of the NLS or from a mean field approximation of the BHH. In the latter case, one can propagate the field operator $\hat{b}_{i}$ forward in time using the BHH in the Heisenberg picture:  $i \hbar \partial{}_t \hat{b}_{i}=[\hat{b}_{i},\hat{H}]$. Assuming the many-body state is a product of Glauber coherent states, $\langle \hat{b}_i^{\dagger}\hat{b}_i\hat{b}_i\rangle = \psi_i^*\psi_i\psi_i$, where $\psi_i\equiv\langle \hat{b}_{i} \rangle$, leads to the DNLS:
\begin{equation}
\label{eq:DNLS}
\textstyle i\hbar\dot{\psi_i}=-J(\psi_{i+1}+\psi_{i-1}) + g|\psi_i|^2 \psi_i+V^{\mathrm{ext}}_{i} \psi_i.
\end{equation}
In Eq.~\eqref{eq:DNLS}, the condensate order parameter, $\psi_i$, is normalized to the number of particles, $N = \sum^{L}_{i=1} |\psi_i|^2$. Mean field simulations are performed using a pseudospectral, fourth-order Runge-Kutta, adaptation of Eq.~\eqref{eq:DNLS}.  The BHH approaches the DNLS in the mean field limit $N\to \infty$, $U \to 0$, $NU/J=\mathrm{const.}$  We emphasize that both the BHH and the DNLS are single band models, valid when the soliton covers many sites; a continuum limit is possible for $\nu J = \mathrm{const.}$, $\nu \equiv N/L \to 0$ and $J \to \infty$; however, a continuum limit would restrict us numerically to very small numbers of particles~\cite{muth2010}.

\begin{figure}[t]
\begin{center}
\epsfxsize=7.5cm \epsfysize=3.0cm \leavevmode \epsfbox{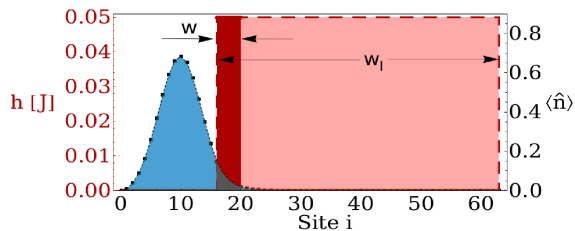}
\caption{\label{fig:InitialState} \emph{Initial State.}  Bright soliton formed by relaxation in imaginary time with a barrier of height $h$ and initial width $w_I$ (dashed line). Before real time propagation the barrier is reduced to width $w$ (solid red line) so the soliton can commence macroscopic quantum tunneling.}
\end{center}
\end{figure}
\begin{figure}[t]
\begin{center}
\epsfxsize=8.6cm \epsfysize=5.5cm \leavevmode \epsfbox{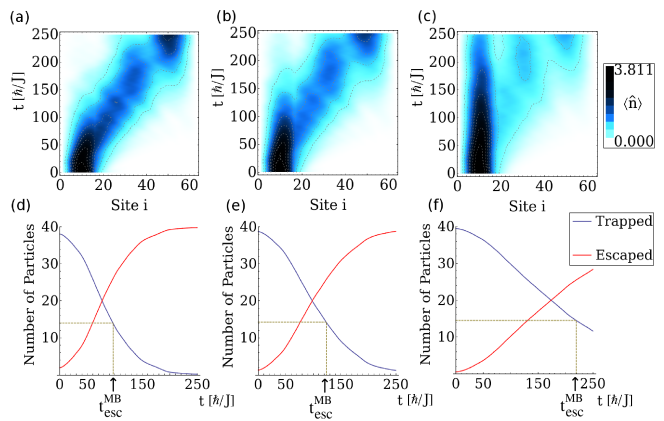}
\caption{\label{fig:AvgNumChangeW} \emph{Many-Body Tunneling, and Calculation of Decay Time.}
Average particle number per site (top row) and number of trapped (blue) and escaped (red) particles (bottom row), for barrier widths $w=1$ (left), $w=2$ (middle), and $w=4$ (right); the $1/e$ decay time is $t_{\mathrm{esc}}^{\mathrm{MB}}=96.9$, $123.9$, and $219.4$, all $\pm 1.25$, respectively.}
\end{center}
\end{figure}

We initialize the many-body wavefunction via imaginary time relaxation as a bright soliton trapped behind the barrier as illustrated in Fig.~\ref{fig:InitialState}. We set $V^{\mathrm{ext}}$ to height $h=0.05$ and width $w_{I}$, effectively reducing the system size. At $t=0$, in real time, the barrier is decreased to width $w$, where $w$ is typically one to five sites, such that the soliton can escape on a time scale within reach of TEBD simulations. Attractive interactions $U<0<h$ ensure that tunneling is always quantum, not classical. We choose $L$ large enough so that reflections from the box boundary at the far right do not return to the barrier in $t < t_{\mathrm{esc}}$. Evolving in real time, we first make a coarse observation of the dynamics of MQT in Fig.~\ref{fig:AvgNumChangeW} by plotting the average particle number in different regions, in order to determine $t_\mathrm{esc}$.

\begin{figure}[t]
\begin{center}
\epsfxsize=8.5cm \epsfysize=7.5cm \leavevmode \epsfbox{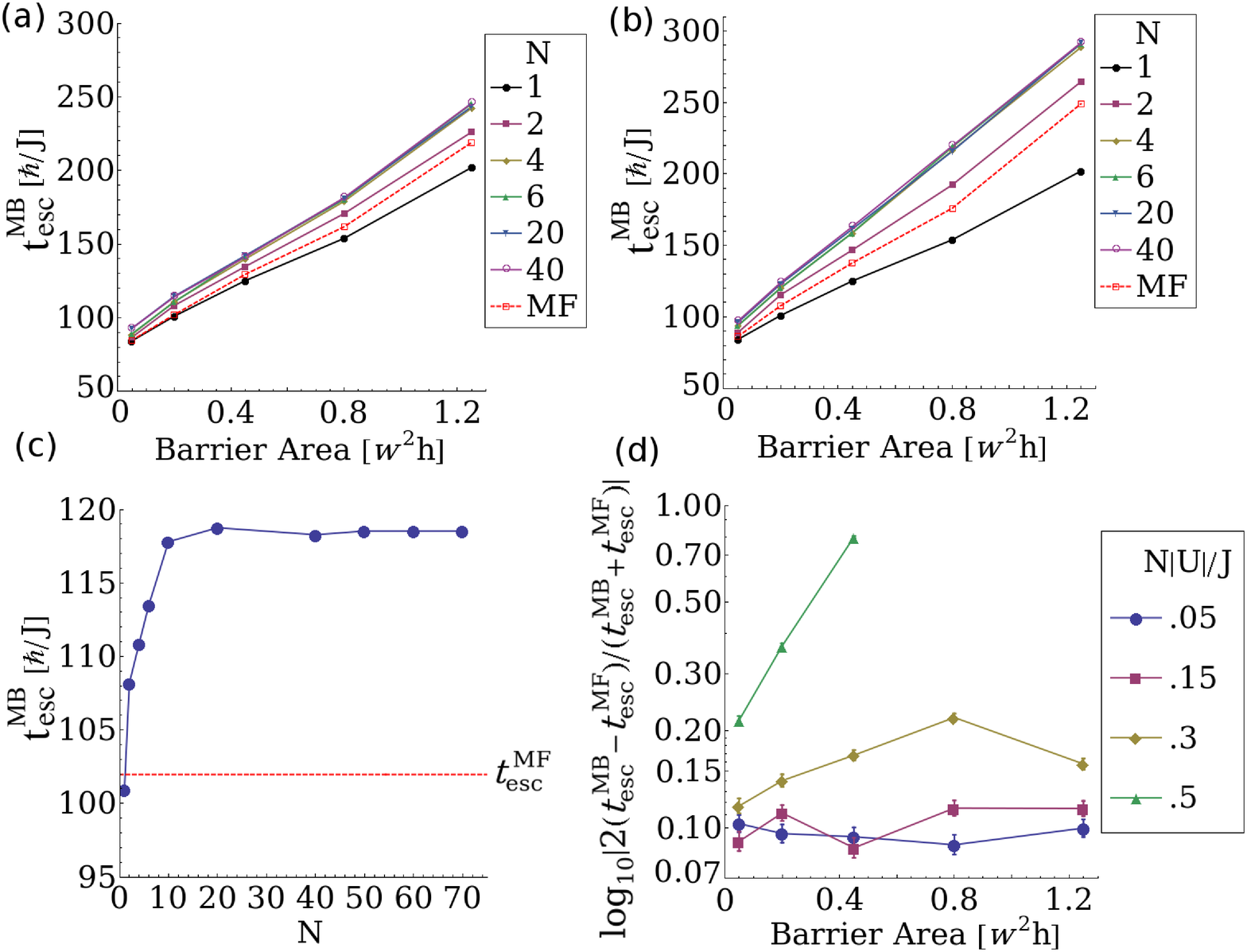}
\caption{\label{fig:NUovJEscapeSummary} \emph{Many Body vs. Mean Field Escape Time Predictions.} (a)-(b) Dependence of $t_{\mathrm{esc}}^{\mathrm{MB}}$ on barrier area and particle number for (a) $N U/J = -0.15$ and (b) $-0.3$. (c) $t_{\mathrm{esc}}^{\mathrm{MB}}$ plateaus for $10$ to $70$ particles as shown for $N U/J = -0.15$. (d) In the plateau region of $N=40$, $t_{\mathrm{esc}}^{\mathrm{MB}}$ significantly differs from $t_{\mathrm{esc}}^{\mathrm{MF}}$ for a range of barrier areas and interaction strengths. Curves are a guide to the eye, points represent actual data.}
\end{center}
\end{figure}

How do many-body predictions compare to mean field ones?  We define $t_{\mathrm{esc}}^{\mathrm{MF}}$ and $t_{\mathrm{esc}}^{\mathrm{MB}}$ as the mean field and many-body escape times, respectively.  For fixed $N U/J$, $w$, and $h$, the DNLS gives the same result independent of $N$ and $U$; $t_{\mathrm{esc}}^{\mathrm{MB}} \to t_{\mathrm{esc}}^{\mathrm{MF}}$ only for $N \to \infty$ and $U \to 0$; and $w^2 h$ determines the barrier area. Figure~\ref{fig:NUovJEscapeSummary} illustrates our exploration of this parameter space.
The dynamics of MQT predicted by the DNLS and BHH differ strongly. Generally, the DNLS grossly under predicts $t_{\mathrm{esc}}$ when $N |U|/J$ is sufficiently large. For example, in Fig.~\ref{fig:NUovJEscapeSummary}(c) for $N|U|/J = 0.15$, except for $N=1$, the BHH predicts an increase in $t_{\mathrm{esc}}$ over the DNLS, approaching a nearly constant value for $\nu\simeq 1/2$ to 4. Although out of range of our simulations, we expect $t_{\mathrm{esc}}^{\mathrm{MB}}$ to subsequently decrease back down to $t_{\mathrm{esc}}^{\mathrm{MF}}$, when $\nu \gg 1$. Escape times follow the same qualitative pattern for higher $N|U|/J$.  The increase is due to number fluctuations, which are not permitted by mean field theory; particles spend more time further from the barrier at the soliton peak, and thus take longer to tunnel.

Systematic error in $t_{\mathrm{esc}}^{\mathrm{MB}}$ results from the Schmidt truncation used in TEBD~\cite{vidal2004}, $\chi$, the truncation in the on-site Hilbert space dimension, $d$, and the time resolution at which we write out data, $\delta t$.  The hardest many-body measures to converge, such as the block entropy, at $\chi=35$ have an error $\lesssim 10^{-3}$ for $N=70$, and were checked up through $\chi=55$; due to small $U$ and effective system size, much lower $\chi$ is required than usual in TEBD.  Error bars in Fig.~\ref{fig:NUovJEscapeSummary}(d) are due solely to $\delta t$; there is additional error from our Suzuki-Trotter expansion which is smaller than $\chi$-induced error.  For up to $N=20$ we have not truncated $d$, but for larger $N$ up to 70, we truncated to $d=18$.  A lower truncation results in decreased $t_{\mathrm{esc}}^{\mathrm{MB}}$, e.g. by 10\% for $d=5$, $NU/J=-0.1$, and $N=10$, even though $\mathrm{max}(\langle\hat{n}\rangle) < 1$, since more weight is given to spread-out Fock states. The attractive BHH requires much higher $d$ than the repulsive BHH, since $U<0$ increases number fluctuations in high density regions, i.e., at the soliton peak.  The BHH also has a number of sources of systematic error, the most important of which is virtual fluctuations to the second band; however, since we compare single-band DNLS to single-band BHH this does not effect our comparison.

\begin{figure}[t]
\begin{center}
\epsfxsize=8.3cm \epsfysize=6.4cm \leavevmode \epsfbox{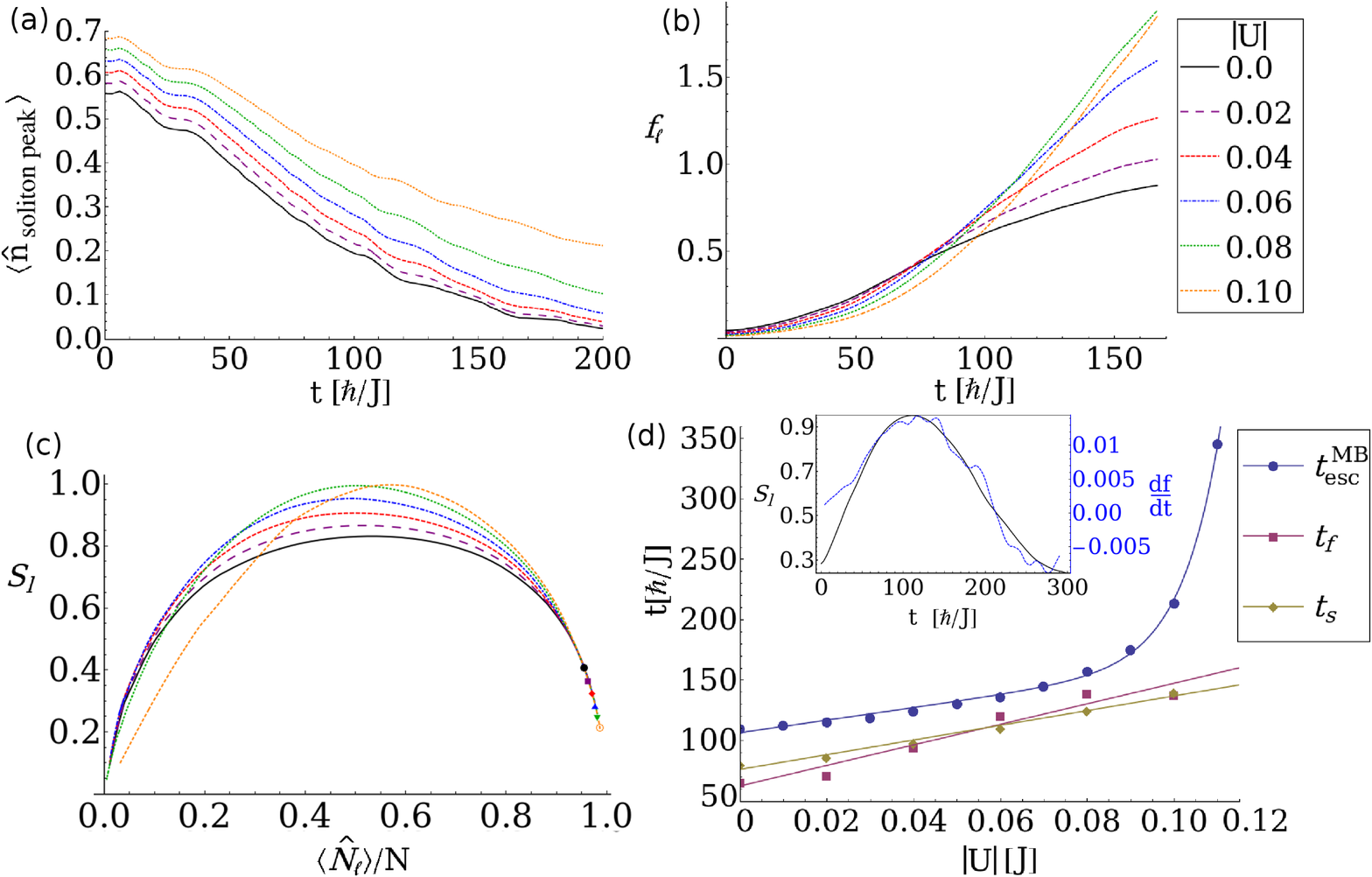}
\caption{\label{fig:peakFluctEnt} \emph{Many-body Quantum Measures.}  (a) Average number at the peak of the soliton shows bursts of particles~\cite{dekel2010}.  (b) Fluctuations in the number of trapped particles increases with $|U|$. (c) Universal curve for the entropy of entanglement vs. the average number of trapped particles. (d) Exponential vs. linear increase in escape vs. many body times as a function of interactions.  (Inset) Block entropy and $df_l/dt$ closely follow each other, here for $|U|=0.06$. The key applies to panels (a)-(c) and all plots treat $N=6$.}
\end{center}
\end{figure}
In Fig.~\ref{fig:peakFluctEnt}(a) we plot the average number at the peak of the soliton. There are points in time when the number density exhibits steep exponential decay, and others during which it is nearly constant, similar to the density bursts found by Dekel \textit{et al.}~\cite{dekel2010}; thus their predictions are correct even in the many-body regime.  The first burst is independent of $U$. The initial flat horizontal region, at $t\simeq 25$, originates from initially left moving particles that are reflected off the leftmost infinite boundary and return back to the barrier. All subsequent deviations from exponential decay appear to be dependent on $U$. In the coarser measure $t_{\mathrm{esc}}^{\mathrm{MB}}$, we find exponential scaling dominates for stronger interactions, as shown in Fig.~\ref{fig:peakFluctEnt}(d).  The dependence is exponential for two reasons: the many-body wavefunction tends to have large number fluctuations at the soliton peak, keeping particles away from the barrier; and the averaged density creates an effective potential which increases the effective barrier size, as in mean field theory.  Results in Fig.~\ref{fig:peakFluctEnt} are for $N=6$; we found qualitatively similar results for up to $N=20$, although simulations are limited in the large $|U|$ regime.

To characterize the quantum nature of MQT, in Fig.~\ref{fig:peakFluctEnt}(b) we plot the fluctuations in the number of particles behind the barrier, $f_l=(\langle N^2_l \rangle - \langle N_l \rangle^2) / \langle N_l \rangle$, where $N_l$ is the number of particles to the left of site $l$ and $l$ is taken at the outer edge of the barrier. Once MQT commences, the maximum value of $f_l$ in time increases with $|U|$ because number densities just outside the barrier have more influence to ``pull'' additional particles through the barrier, and vice versa. For early times, less than $t=50$ in Fig.~\ref{fig:peakFluctEnt}(b), $f_l$ increases faster for smaller values of $|U|$ because the initial soliton is wider, but interactions take over shortly thereafter.

Song \emph{et al.}~\cite{song2010} have shown that in 1D conformal systems for which there is a conserved quantity, such as particle number, the variance of the fluctuations in that quantity between two subsystems $A,B$ scales with the von Neumann entanglement entropy between $A,B$. Of particular interest to MQT is the von Neumann block entropy characterizing entanglement between the remaining particles and the escaped particles, $S_l\equiv -\mathrm{Tr}(\hat{\rho}_l \log \hat{\rho}_l)$, where $\hat{\rho}_l$ is the reduced density matrix for the well plus barrier.  The key features of $S_l$ are illustrated in a universal curve in Fig.~\ref{fig:peakFluctEnt}(c): on the lower right side tunneling has not yet commenced; $S_l$ maximizes part way through the tunneling process in the center of the curve; and $S_l$ then decreases again to the left as the particles finish tunneling out. Defining $t_s$ as the time at which $S_l$ is maximized and $t_f$ as the time at which the slope of the number fluctuations, $df_l/dt$, is maximized, we find $t_s$ and $t_f$ both increase linearly with $U$, in contrast to $t_\mathrm{esc}^{\mathrm{MB}}$, as shown in Fig.~\ref{fig:peakFluctEnt}(d). Moreover, $S_l$ and $df_l/dt$ follow the same general trends in time, as shown in the inset of Fig.~\ref{fig:peakFluctEnt}(d). The two do not scale precisely, as $df_l/dt$ is distorted by the density bursts illustrated in Fig.~\ref{fig:peakFluctEnt}(a).

\begin{figure}[t]
\begin{center}
\epsfxsize=8.3cm \epsfysize=6.2cm \leavevmode \epsfbox{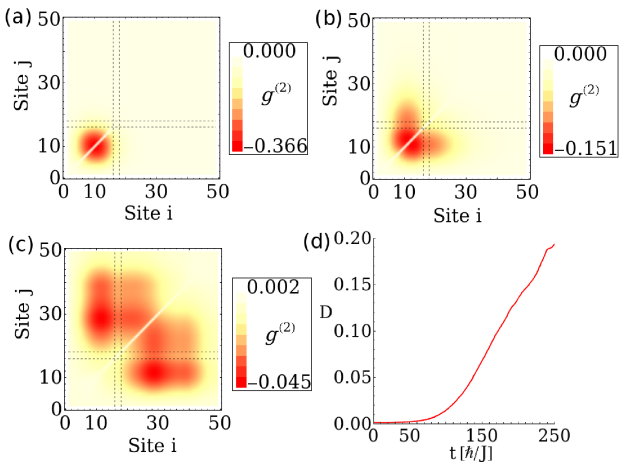}
\caption{\label{fig:geetwo} \emph{Time-dependence of Density-density Correlations.}  (a)-(c) $g^{(2)}$ shows correlations between trapped and escaped particles. The barrier, indicated by dashed lines, breaks up negatively-correlated regions (red); shown are time slices at $t=0,62,125$. (d) Quantum depletion also grows rapidly.}
\end{center}
\end{figure}

Another experimental signature is the density-density correlations, $g^{(2)}_{ij}=\langle \hat{n}_i \hat{n}_j \rangle - \langle \hat{n}_i \rangle \langle \hat{n}_j \rangle$, extractable from noise measurements~\cite{altman2004}; $g^{(2)}$ is zero in mean field theory.  As customary, we subtract off the large diagonal matrix elements of $g^{(2)}$ to view the underlying off-diagonal structure.  In Fig.~\ref{fig:geetwo}(a)-(c) we show $g^{(2)}$ for $N=40$, $NU/J=-0.015$, and $w=2$, dividing up the system to observe correlations between the three physical regions: trapped, under the barrier, and escaped.  We initially observe near-zero correlations everywhere except near the soliton peak. At $t_s$, $g^{(2)}$ shows many negatively-correlated regions ($g^{(2)}<0$) which are broken up by the potential barrier. In Fig.~\ref{fig:geetwo}(d) we also show, for $N=20$, rapidly growing quantum depletion $D \equiv 1 - (\sum_{m=2}^{L}\lambda_m) / (\sum_{m=1}^{L}\lambda_m)$ where $\{\lambda_m\}$ are the eigenvalues of the single particle density matrix $\langle \hat{b}_i^{\dagger}\hat{b}_j\rangle$.  This growth in $D$ emphasizes the many-body nature of the escape process.

In conclusion, we have performed quantum many-body simulations of the macroscopic quantum tunneling of bright solitons using TEBD to time-evolve the attractive Bose-Hubbard Hamiltonian. We found strong deviations from mean field predictions.  The escape time was shown to increase exponentially with interactions while both block entropy and the slope of number fluctuations maximized at a time which scaled linearly; entropy generally followed closely the slope of number fluctuations, suggesting a dynamical extension of the static concepts of Song \textit{et al.}~\cite{song2010}.  Our study shows that many-body effects in macroscopic quantum tunneling can be observed via number fluctuations and density-density correlations as well as the increased escape time.

We thank Veronica Ahufinger, Mark Lusk, Jen Glick, Ana Sanpera, Michael Wall, and David Wood for valuable discussions. This work was supported by NSF.


\end{document}